\documentclass[aps,pre,preprint,superscriptaddress]{revtex4-1}

\usepackage{graphicx, amssymb, verbatim, amsmath, siunitx, color, float, soul, makecell, xcolor}
\usepackage[normalem]{ulem}

\usepackage[justification=raggedright, singlelinecheck=false]{caption}
\usepackage{natbib,url, enumerate}
\usepackage[colorlinks, allcolors = blue, bookmarksopen,bookmarksnumbered]{hyperref}

\definecolor{colsq}{rgb}{0,0.4470,0.7410}
\definecolor{coltr}{rgb}{0.8500,0.3250,0.0980}
\definecolor{coldm}{rgb}{0,0.4470,0.7410}
\definecolor{colbtr}{rgb}{0.9290,0.6940,0.1250}

\begin{document}

\title{Violation of the Fluctuation Dissipation Theorem during Domain Growth in the Long-range Ising Model}

\author{Parbati Saha}
\email{parbati.saha@physics.iitd.ac.in}
\affiliation{Department of Physics, Indian Institute of Technology Delhi, New Delhi -- 110016, India.}
    
\author{Sanjay Puri}
\email{purijnu@gmail.com}
\affiliation{School of Physical Sciences, Jawaharlal Nehru University, New Delhi -- 110067, India.}
    
\author{Varsha Banerjee}
\email{varsha@physics.iitd.ac.in}
\affiliation{Department of Physics, Indian Institute of Technology Delhi, New Delhi -- 110016, India.}

\begin{abstract}
The celebrated {\it fluctuation dissipation theorem} (FDT) does not apply to nonequilibrium systems. In this context, Cugliandolo and Kurchan [Phys. Rev. Lett. {\bf 71}, 173 (1993)] introduced a generalized FDT which interprets the nonequilibrium evolution as a composition of {\it time sectors} corresponding to different {\it effective temperatures}. We use this framework to study domain growth in the $d=2$ long-range Ising model (LRIM) with nonconserved kinetics ({\it Glauber spin-flip}) and conserved kinetics ({\it Kawasaki spin-exchange}). We study the dynamical scaling and super-universal (SU) scaling of various two-time quantities, e.g., autocorrelation function, response function, effective temperature, etc. In particular, we investigate how the interaction range and conservation laws affect these characteristic features of domain growth.
\end{abstract}

\maketitle

\section{Introduction}
\label{s1}

An important and well-studied nonequilibrium phenomenon is the {\it kinetics of phase transitions} \cite{Puri2009,Bray2002}. This refers to the evolution of a system after it is rendered thermodynamically unstable by (say) a rapid quench at time $t=0$ from a high-temperature disordered state to a low-temperature ordered state. Such a system does not order instantaneously. Instead, the system evolves via the emergence and growth of domains that are separated by interfaces or defects. In many simple systems, this coarsening kinetics is characterized by a power-law growth: $\ell(t) \sim t^{1/z}$, where $\ell(t)$ is the typical length scale and $z$ is the dynamical exponent. The determination of domain growth laws is an important aspect of coarsening experiments or simulations, as it reveals details of the free-energy landscape and relaxation time scales in the system. In theoretical studies, coarsening systems are often described by Ising models with stochastic kinetics induced by a heat bath \cite{dp04}. There is a good understanding of growth laws in disorder-free and isotropic Ising models with nearest-neighbor (NN) interactions. For example, systems with nonconserved spin-flip Glauber dynamics (GD) obey the Lifshitz-Allen-Cahn (LAC) law: $\ell(t)\sim t^{1/2}$. On the other hand, systems with conserved spin-exchange Kawasaki dynamics (KD) follow the Lifshitz-Slyozov law: $\ell(t)\sim t^{1/3}$. This power-law growth is characteristic of systems without $\ell$-dependent energy barriers and a unique relaxation time-scale. Barriers dependent on $\ell$ are characteristic of spin systems with quenched disorder and frustration.

The nonequilibrium evolution in disordered spin systems (e.g., spin glasses, random bond Ising model or RBIM, random field Ising model or RFIM) is somewhat different; they usually show logarithmic growth at late times \cite{lippiello2010,corberi2011,corberi2012}. This is a consequence of the slow relaxation that arises from a complex free-energy landscape with deep local minima separated by high-energy barriers, leading to a plethora of relaxation times. Concomitant with slow relaxation is the phenomenon of {\it aging} \cite{Henkel2008}. As the system ages, it faces larger barriers because it has already crossed the smaller barriers. An older system therefore experiences progressively deeper local minima as the system has more time to explore the landscape. Thus, the evolution becomes slower or more sluggish with time -- hence the term {\it aging}. This property is studied using two-time quantities \cite{Henkel2008,mz09}, such as the {\it autocorrelation function}, which relates a local spin variable $s(t)$ at different times:
\begin{equation}
    C(t,t_{w}) = \langle s(t)s(t_{w}) \rangle - \langle s(t) \rangle \langle  s(t_{w}) \rangle , \quad t > t_w,
    \label{Autocor}
\end{equation}
where $t_w$ is the {\it waiting time} or {\it annealing time}. The angular brackets in Eq.~(\ref{Autocor}) denote an average over independent runs, obtained from distinct initial conditions and noise realizations. Unlike systems in equilibrium, $C(t,t_w)$  does not exhibit time-translation invariance; rather, the time scale for the autocorrelation decay increases with $t_w$. Typically, $C(t,t_w)$ is described by a universal scaling function $f(x)$ such that
\begin{equation}
C(t,t_{w}) = f\left( \frac{\ell(t)}{\ell(t_w)} \right) .
\end{equation}

In many applications, one is interested in the response of an aging system to an external perturbation. Consider an external field $h(t_w)$ applied to the spin system in the window $\left[t_w,t_w + \Delta t \right]$. The {\it linear response function} is defined as follows:
\begin{equation}
    R(t,t_w) = \lim_{\Delta t \rightarrow 0}\frac{1}{\Delta t}\frac{\partial \langle s(t)\rangle}{\partial h(t_w)} \bigg|_{h=0}.
\label{LRFn}    
\end{equation}
In equilibrium, two-time quantities satisfy translational invariance in time, that is, $C(t,t_{w})=C(t-t_{w})$ and $R(t,t_w)=R(t-t_w)$. The {\it fluctuation dissipation theorem} (FDT) holds in equilibrium and relates the correlation and response functions:  
\begin{equation}
R(t-t_w) = \frac{1}{T}\frac{\partial}{\partial t_w}C(t-t_w), 
\label{FDT}
\end{equation}
where $T$ is the equilibrium temperature of the bath. Thus, the FDT constitutes a relationship between fluctuations (random motion of the system due to thermal noise) and response (the reaction of the system to external perturbations).

As mentioned above, aging systems lose translational symmetry in time and do not obey the FDT. In this context, Cugliandolo and Kurchan \cite{cugliandoloKurchan1993,cugliandolokurchan1994} proposed the notion of {\it fluctuation dissipation ratio} (FDR), as a measure of the deviation from equilibrium:
\begin{equation}
X(t,t_w) = \frac{T R(t,t_w)}{\partial_{t_{w}}C(t,t_{w})}.
\label{FDR}
\end{equation}
Clearly, systems in equilibrium are characterized by $X(t,t_{w}) = 1$, while nonequilibrium behavior yields $X(t,t_{w}) \neq 1$. Cugliandolo and Kurchan also introduced the concept of effective temperature $T_{\rm eff}$ by rewriting Eq.~(\ref{FDR}) to yield a {\it generalized} FDT:
\begin{equation}
R(t,t_w) = \frac{X(t,t_{w})}{T}\frac{\partial}{\partial t_w}C(t,t_w) \equiv \frac{1}{T_{\rm eff}}\frac{\partial}{\partial t_w}C(t,t_w).
\label{GFDT}
\end{equation}
For systems in equilibrium, $T_{\rm eff}=T$. As emphasized earlier, in slowly relaxing systems, different degrees of freedom have different time scales. The nonequilibrium evolution can thus be interpreted as a sequence of time sectors corresponding to distinct effective temperatures.

In experiments, the applied field is not an impulse but acts over an interval of time. Consequently, experimental observations are better represented by the so-called {\it integrated response function} (IRF), which is the cumulative response of the system to a field applied over a time window $[t_w,t-\epsilon]$ with $t_w\le t-\epsilon$ \cite{Lippiello2005}:
\begin{equation}
    \chi(t,[t_{w},t-\epsilon]) = \int_{t_w}^{t-\epsilon} dt^{\prime}R(t,t^{\prime}) = \int_{t_{w}}^{t-\epsilon} dt^{\prime}\frac{X(t,t^{\prime})}{T} \frac{\partial}{\partial t^{\prime}}C(t,t^{\prime}).
\label{IntRes1}  
\end{equation}
Since $\epsilon$ is generally taken to be the smallest time interval used in the calculation, it is reasonable to designate $\chi(t,[t_{w},t-\epsilon]) \simeq \chi(t,t_{w})$. In addition, for systems where dynamics is slow, it is reasonable to assume that $X(t, t')$ varies slowly in the integration window, so that $X(t,t^{\prime}) \simeq X(t,t_{w})$ for all $t^{\prime} \in [t_{w},t]$. This yields a simplified form for the IRF:    
\begin{equation}
\chi(t, t_w) \simeq \frac{X(t, t_w)}{T} \int_{t_w}^{t-\epsilon} dt' \, \frac{\partial C(t, t')}{\partial t'} = \frac{X(t, t_w)}{T}[C(t, t - \epsilon) - C(t, t_w)].
\label{IRF}
\end{equation}
Now, because $ \epsilon$ is infinitesimal, we take $C(t, t - \epsilon) \simeq C(t, t)$. For an Ising spin variable, $C(t,t)=1$ and Eq.~(\ref{IRF}) further simplifies as
\begin{equation}
\chi(t, t_w) \simeq \frac{X(t, t_w)}{T} [1 - C(t, t_w)] = \frac{1-C(t,t_w)}{T_{\rm eff}}.
\label{IRFT_eff}
\end{equation}
In equilibrium, $T_{\rm eff} = T$ and Eq.~(\ref{IRFT_eff}) reduces to a linear relation referred to as the FDT line:
\begin{equation}
T\chi(t,t_{w}) = 1-C(t,t_{w}).
\label{FDT_line}    
\end{equation} 
Deviations from this behavior indicate a violation of the FDT, as the system is not in equilibrium \cite{Barrat1998, ricci2003, chatelain2003, godreche2004, Krzakala2005}. In slowly relaxing systems, the plot of $T\chi(t,t_{w})$ vs. $C(t,t_{w})$ is often used to identify the FDR, as well as the time sectors corresponding to different values of $T_{\rm eff}$.

The preceding discussion may suggest that aging is characteristic of systems with complex free-energy landscapes that arise from disorder and frustration. However, the Ising model (IM) without disorder or frustration also exhibits aging during coarsening, making it the simplest system to study the above quantities. Several insightful studies of aging during domain growth have been performed in the NN Ising model (IM) without disorder \cite{mz09}. However, spin-spin interactions in most magnetic materials extend beyond the NNs to large distances. Therefore, a more realistic picture requires the use of models that incorporate long-range interactions. Although the relevance of such models has been evident for a long time, their study presents major analytical and computational difficulties. As a result, the long-range Ising model (LRIM), though of immense experimental importance, continues to be less explored. In this paper, we conduct a detailed study of aging in the $d=2$ LRIM.

In recent years, considerable study of domain growth in the LRIM has been performed. Several authors have introduced efficient Monte Carlo (MC) procedures \cite{Newman1999} that allow exploration of the LRIM for large system sizes and long time scales. These have confirmed analytical predictions \cite{Bray1994} for growth laws and spatial scaling properties for both GD and KD \cite{christiansen2019,christiansen2020, Agrawal2021,Fabio2022,Fabio2024,Subir2024}. However, there has only been a restricted study of the two-time quantities discussed above, with a focus on the decay of $C(t,t_w)$. In this paper, we address this shortcoming. In particular, we examine the violation of the FDT during domain growth in the LRIM by evaluating the IRF and $T_{\rm eff}$.

This paper examines how aging in the LRIM is affected by (a) the interaction range and (b) the presence of a conservation law. Section~\ref{s2} provides the details of the LRIM and the theoretical framework required for our far-from-equilibrium study. Detailed numerical results are presented in Sec.~\ref{s3}. The paper ends with a summary and discussion of our results in Sec.~\ref{s4}.

\section{Theoretical Framework}
\label{s2}

The Hamiltonian of the $d=2$ LRIM with $N=L^2$ spins is given by 
\begin{equation}
\mathcal{H} = - \sum_{i<j}J_{ij}s_{i}s_{j}, \quad s_i = \pm 1,
\end{equation}
where
\begin{equation}
J_{ij} = \frac{J_0}{r_{ij}^{2+\sigma}} .
\end{equation}
Here, $J_0$ sets the scale of the exchange interaction, and $r_{ij} = |\vec{r}_i - \vec{r}_j|$ is the distance between spins at the sites $i$ and $j$. The exponent $\sigma$, which regulates power-law decay, smoothly transitions from NN interactions ($\sigma = \infty$) through intermediate ranges ($\sigma \simeq 1$) to long-range interactions ($\sigma < 1$). This system exhibits a phase transition from a paramagnetic state to a ferromagnetic state at the critical temperature $T_c(\sigma)$ \cite{Horita2017}. 
 
As Ising spins do not have intrinsic dynamics, the system is placed in contact with a heat bath at temperature $T$ that generates stochastic spin moves \cite{dp04}. For GD, we consider spin flips where a randomly chosen spin $s_m$ is flipped to $-s_m$. The acceptance probability for a flip is based on the standard Metropolis criterion \cite{metropolis1953}: 
\begin{equation}
p = \mbox{min}[1,\exp(-\beta \Delta E)] , 
\end{equation}
where $\beta= (k_BT)^{-1}$ and $\Delta E = {\cal H}^{\rm new} - {\cal H}^{\rm old}$ is the energy change due to the flip. Since interactions are LR, each Metropolis step involves computing $O(N)$ sums to evaluate $\Delta E$, making these evaluations computationally demanding. For efficiency, it is useful to store the effective field $h_{i} = \sum_{j \neq i}J_{ij}s_{j}$ at each site at the beginning of the simulation. An update of $\{h_i\}$ is then required only if a spin-flip is accepted. For a spin flip $s_m \rightarrow -s_m$, we have the following:
\begin{equation}
\Delta E = 2 s_m h_m .
\label{glauber}
\end{equation}
If this flip is accepted, the effective fields are updated as
\begin{eqnarray}
h_i^{\rm new} &=& h_i - 2 J_{mi} s_m, \quad i \neq m , \nonumber \\
h_m^{\rm new} &=& h_m .
\end{eqnarray}

Let us next discuss KD, where the conservation law is implemented by considering stochastic exchanges ($s_n \leftrightarrow s_m$) of NN spins. The corresponding energy change is
\begin{equation}
\Delta E = (s_m-s_n)(h_m-h_n) + J_{mn} (s_m-s_n)^2 .
\label{kawasaki}
\end{equation}
If this exchange is accepted, the update of effective fields is as follows:
\begin{eqnarray}
h_i^{\rm new} &=& h_i - (J_{mi}-J_{ni}) (s_m-s_n), \quad i \neq m,n , \nonumber \\
h_m^{\rm new} &=& h_m + J_{mn} (s_m-s_n) , \nonumber \\
h_n^{\rm new} &=& h_n - J_{mn} (s_m-s_n) .
\end{eqnarray}
The effective field approach significantly speeds up the computation at low temperatures, where many moves are rejected. This method can accelerate the simulation by up to a factor of $10^{3}$ compared to the standard Metropolis algorithm.

Simulations involving LR interactions exhibit strong finite-size effects for small values of $\sigma$. Much larger lattices are required as compared to the case with short-range interactions. These can be created by employing periodic boundary conditions with the minimum-image convention. The interaction strength between spins in a square lattice gradually decreases until it reaches a specific cutoff distance. The effective interaction strength can be written as an infinite sum over all images:
\begin{equation}
J_{ij} = J_0 \sum_{\vec{n}} \frac{1}{|\vec{n} + \vec{r}_{i}- \vec{r}_{j}|^{2+\sigma}} ,
\end{equation}
where $\vec{n} = (n_{x},n_{y},...)$ is the displacement vector with $n_{\alpha} = 0, \pm L, \pm 2L,....$ denoting the coordinates of imaginary copies. Infinite sums are computed using the Ewald summation technique, which is a standard method to deal with LR interactions \cite{Frenkel1996}.

Let us now consider a quench of the LRIM from a high-temperature paramagnetic state ($T>T_c(\sigma)$) to the ferromagnetic phase ($T< T_{c}(\sigma)$). The system develops correlated domains (regions enriched in up- or down-spins) that coarsen with time. The usual probe to quantify the morphologies is the spatial correlation function:
\begin{equation}
    C_{\rm sp}(\vec{r},t) = \frac{1}{N}\sum_{i=1}^{N} \left[\langle s_i(t) s_j(t) \rangle - \langle s_i(t) \rangle \langle s_j(t) \rangle \right] ,
    \label{Cor1}
\end{equation}
where $\vec{r} = \vec{r}_i - \vec{r}_j$. (Here, we have assumed that the system is translationally invariant.) The characteristic length scale $\ell(t)$ is defined as the distance at which $C_{\rm sp}(\vec{r},t)$ decays to some fraction of its maximum value. If domain growth is characterized by a unique length scale $\ell(t)$, the morphology is unchanged with time, except for a scale factor. In this case, the correlation function demonstrates a dynamical scaling property: $C_{\rm sp}(\vec{r},t) = f(r/ \ell )$, where $f(x)$ is the scaling function \cite{Puri2009}.

In an important paper, Bray and Rutenberg (BR) \cite{Bray1994} predicted growth laws in the LRIM using an energy scaling argument. For systems with GD (nonconserved dynamics), they obtained
\begin{equation}
    \ell(t) \sim t^{1/z}=
    \begin{cases}
        t^{1/(1+\sigma)}, & \sigma < 1,\\
        (t\ln t)^{1/2}, & \sigma = 1, \\
        t^{1/2}, & \sigma > 1.
    \end{cases}
    \label{GL_G}
\end{equation}
For systems with KD (conserved dynamics), BR found that
\begin{equation}
    \ell(t) \sim t^{1/z}=
    \begin{cases}
        t^{1/(2+\sigma)}, & \sigma < 1,\\
        (t\ln t)^{1/3}, & \sigma = 1, \\
        t^{1/3}, & \sigma > 1.
    \end{cases}
    \label{GL_K}
\end{equation}
Eqs.~(\ref{GL_G}) and (\ref{GL_K}) show that growth exponents depend on $\sigma$ in the regime where the effects of LR interactions dominate ($\sigma<1$). 

As mentioned earlier, in addition to spatial correlation functions and growth laws, coarsening systems are also characterized by evaluating two-time quantities. In this context, a standard quantity is the autocorrelation function in Eq.~(\ref{Autocor}). For the kinetic Ising model,
\begin{equation}
C(t,t_{w}) = \frac{1}{N}\sum_{i=1}^{N} \left[ \langle s_{i}(t)s_{i}(t_{w}) \rangle - \langle s_{i}(t) \rangle \langle s_{i}(t_{w}) \rangle \right], \quad t > t_w .  
\label{AutoCor_N}
\end{equation}
This quantity consists of equilibrium and aging parts \cite{mz09}:
\begin{equation}
    C(t,t_{w}) = C_{\rm eq}(t-t_{w}) + C_{\rm ag}(t,t_{w}).
\label{AC_Ag}    
\end{equation}
The contribution $C_{\rm eq}(t-t_{w})$ arises from spins in the interior of the equilibrated domains. It exhibits time-translational invariance and contributes significantly when $t-t_w \ll t_w$. The aging contribution $C_{\rm ag}(t,t_{w})$ arises from the nonequilibrium motion of the interfaces and contributes when $t-t_w \gg t_w$. As expected, it lacks time-translational invariance. For domain growth, the scaling behavior of the aging part is
\begin{equation}
C_{\rm ag}(t,t_{w}) = \tilde{h} \left(\frac{\ell (t)}{\ell (t_{w})}\right) = h \left(\frac{t}{t_{w}}\right) ,
\end{equation}
where $h(x)$ is a scaling function. The latter equation only applies for power-law growth, which is the case in this paper. The decay of $C(t,t_w)$ is determined by the autocorrelation exponent $\lambda$ \cite{fh88}. In the asymptotic scaling regime, we expect 
\begin{eqnarray}
C(t,t_{w}) &\sim & \left(\frac{\ell (t)}{\ell (t_{w})}\right)^{-\lambda} \label{autodecay1} \\
&\sim & \left( \frac{t}{t_{w}} \right)^{-\lambda/z} .
\label{autodecay}
\end{eqnarray}
The presence of quenched disorder generally produces logarithmic growth \cite{lippiello2010,corberi2011,corberi2012}. In that case, we expect the scaling behavior in Eq.~(\ref{autodecay1}) to apply. Yeung et al. \cite{Yeung1996} have used general arguments to obtain lower bounds on $\lambda$. For $d$-dimensional systems with nonconserved kinetics, $\lambda \geq d/2$. On the other hand, for systems with conserved kinetics, $\lambda \geq (2d+1)/2$.

The main focus of this paper is the evaluation of the IRF in Eq.~(\ref{IntRes1}) for the LRIM. From an off-equilibrium generalization of the FDT, Lippiello et al. \cite{Lippiello2005} proposed an efficient numerical algorithm to calculate the response function without the application of a perturbing field. We summarize their results in the following discussion. In the usual framework, the linear response at site $i$ due to the application of a field $h_j$ (at site $j$) for an instant at $t_w$ can be computed from Eq.~(\ref{LRFn}) as
\begin{equation}
    R_{i,j}(t,t_w) = \lim_{\Delta t \rightarrow 0}\frac{1}{\Delta t}\frac{\partial \langle s_{i}(t)\rangle}{\partial h_{j}(t_w)} \bigg|_{h_j=0}.
\end{equation}
Consider next a continued application of the field $h_{j}$ in the time interval $[t_w,\bar{t}~]$ such that $t_w \le \bar{t}~ \le t-\epsilon$, and $\epsilon$ is arbitrarily small. The corresponding contribution to the IRF can be numerically evaluated as \cite{Lippiello2005}
\begin{equation}
   \chi_{i,j}(t,[t_{w},\bar{t}~]) = \epsilon\sum^{\bar{t}~}_{t'=t_{w}}R_{i,j}(t,t').
\end{equation}
The value of $\epsilon=1/N$, which corresponds to a single spin update in a MC step. The summation is performed over discrete times (spaced $\epsilon$ apart) in the interval $[t_{w},\bar{t}~]$. Lippiello et al. \cite{Lippiello2005} show that the above evaluation yields 
\begin{equation}
   \chi_{i,j}(t,[t_{w},\bar{t}~]) 
      = \frac{1}{2 T}\left[C_{i,j}(t,\bar{t}~)-C_{i,j}(t,t_{w})\right] -\frac{\epsilon}{2 T} \sum^{\bar{t}~}_{t'=t_{w}} \langle s_{i}(t-\epsilon)B_{j}(t') \rangle .
 \label{RF_ij}
\end{equation}
Here,
\begin{equation}
    C_{i,j}(t,t') = \langle s_i(t) s_j(t') \rangle - \langle s_i(t) \rangle \langle s_j(t') \rangle 
\end{equation}
and
\begin{equation}
B_{j} = -\sum_{\{s'\}}(s_{j} - s'_{j})w^0(\{s\}\rightarrow \{s'\}) ,
\label{prob}
\end{equation}
where $w^0(\{s\}\rightarrow \{s'\})$ is the transition probability for $\{s\}\rightarrow \{s'\}$ with $h_j = 0$. The transition probability obeys the detailed balance condition \cite{Puri2009}. As usual, the angular brackets indicate an average over independent initial conditions and noise realizations. We focus on the auto-response function with $j = i$. As the system is translationally invariant, the IRF is computed as an average over diagonal terms \cite{Lippiello2005}. We will subsequently present results for the quantity
 \begin{equation}
    \chi(t,t_{w}) =  \frac{1}{N}\sum_{i=1}^N \chi_{i,i}(t,[t_{w},t]).
\label{IRF_av}  
\end{equation}

Eq.~(\ref{RF_ij}) is valid for both GD and KD, although the transition rates used in $B_j(t^{\prime})$ are different in both cases as they depend on $\Delta E$; see Eqs.~(\ref{glauber}) and (\ref{kawasaki}).

\section{Numerical Results}
\label{s3}

We performed MC simulations of the kinetic $d=2$ LRIM in the canonical ensemble (NVT). The use of the Ewald summation technique, along with periodic repetition of the simulation cell, minimized the inaccuracies resulting from the truncation of the interaction range. In addition, the transition probability was obtained at each time increase $\epsilon=1/N$ to calculate the IRF from Eqs.~(\ref{RF_ij})-(\ref{prob}). The system size for the LRIM with GD is $L = 512$, and that with KD is $L = 256$. All statistical data were averaged over at least 20 independent runs, each with different initial conditions and noise realizations.

Simulations were performed for a range of $\sigma$-values. To study domain growth, we performed quenches from $T>T_c (\sigma)$ to $T<T_c (\sigma)$ at time $t=0$. The initial configuration of the lattice $\{s_i(t=0)\}$ consisted of a random distribution of spins with $S_i = +1$ or $-1$, mimicking the disordered state before the quench. The critical temperature (measured in units of $J_0$) depends on the LR interaction parameter, e.g., $T_c=12.555$ for $\sigma=0.6$; $T_c = 5.293$ for $\sigma=1.6$ \cite{Horita2017}; and $T_c = 2.269$ for $\sigma=\infty$. The corresponding quench temperatures for all $\sigma$ were $T = 0.3~T_c$.

\begin{figure}[htb]
\centering
\includegraphics[width=0.9\linewidth]{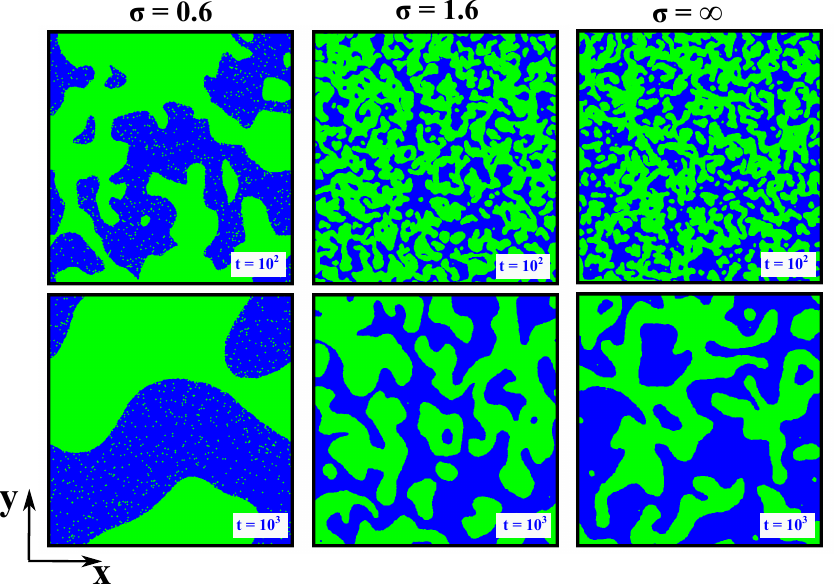}
\caption{Evolution snapshots for the LRIM with GD after a quench at $t=0$. The values of $\sigma$ and $t$ are as shown. The lateral system size is $L = 1024$. The up and down spins are marked blue and green, respectively.}
\label{f1}
\end{figure}

\begin{figure}[htb]
\centering
\includegraphics[width=0.9\linewidth]{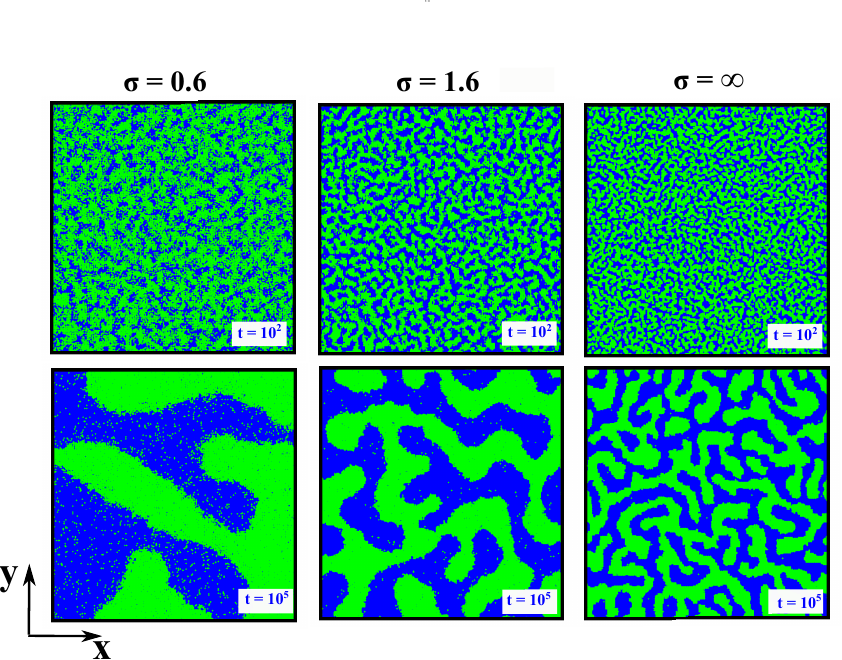}
\caption{Evolution snapshots for the LRIM with KD. The values of $\sigma$ and $t$ are as shown. The lateral system size is $L = 256$. The up and down spins are marked blue and green, respectively.}
\label{f2}
\end{figure}

Figs.~\ref{f1} and \ref{f2} show evolution snapshots for domain growth with GD and KD, respectively. The values of $\sigma$ and $t$ are as specified. As a rule of thumb, the system shows finite-size effects when the domain scale is a significant fraction (say 0.2-0.3) of the lateral system size. The growth is faster for GD than for KD, as expected from the conservation constraint in the latter. In addition, domain growth is faster for smaller values of $\sigma$, as spin exchange interactions are felt up to larger distances.

\begin{figure}[htb]
\centering
\includegraphics[width=0.9\linewidth]{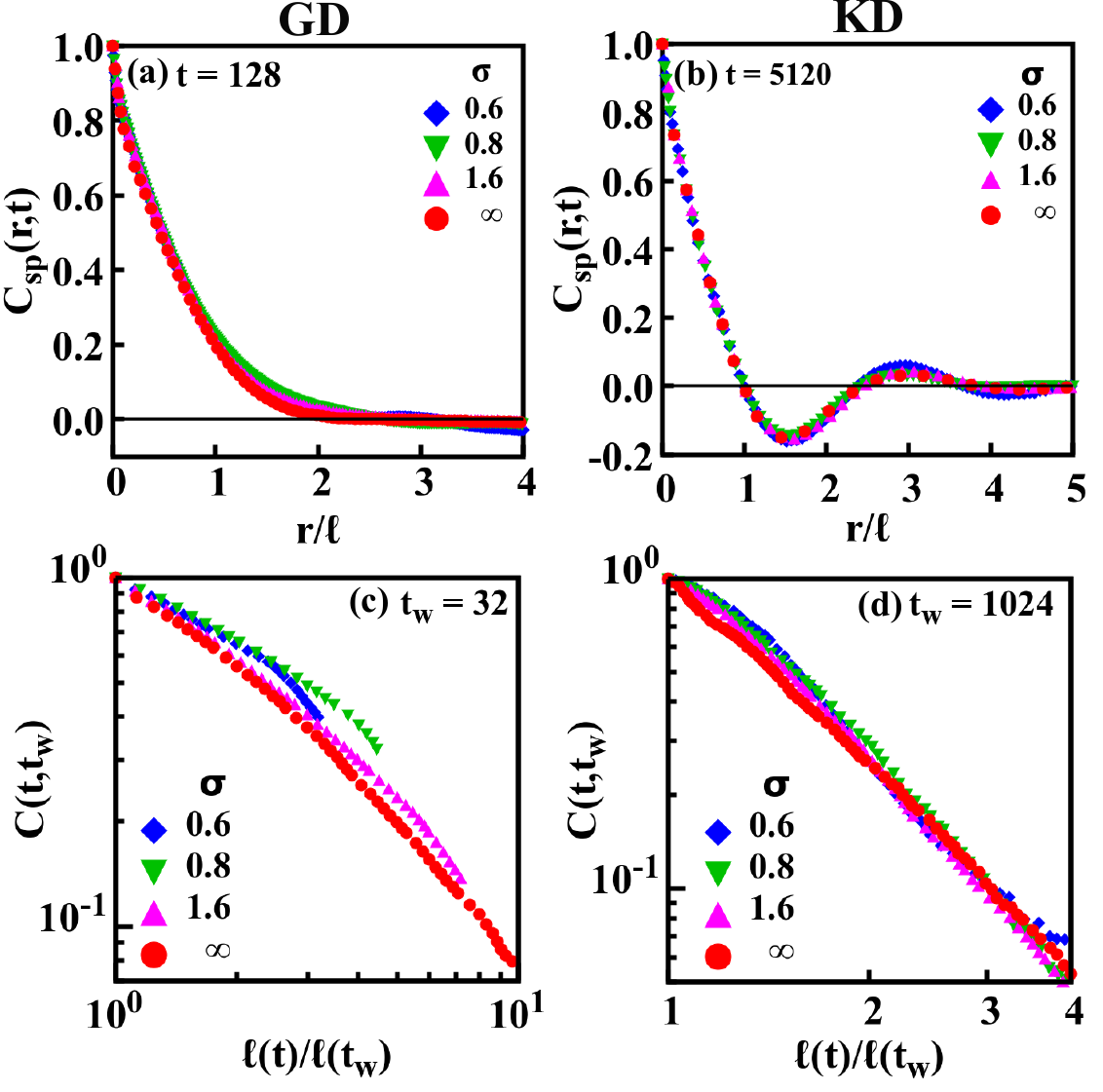}
\caption{The upper row plots $C_{\rm sp}(r,t)$ vs. $r/\ell(t)$ for (a) GD at $t=128$, (b) KD at $t = 5120$. The lower row plots $C(t,t_w)$ vs. $\ell(t)/\ell(t_w)$ on a log-log scale for (c) GD with $t_w=32$, (d) KD with $t_w = 1024$. In all frames, we superpose data for $\sigma = 0.6,0.8,1.6,\infty$.}    
\label{f3}
\end{figure}

Although spatial and auto correlation functions have been well studied in the literature for both GD and KD \cite{christiansen2019, Agrawal2021, Fabio2022, Subir2024}, we present them in Fig.~\ref{f3} for completeness. The upper row shows the scaled correlation function $C_{\rm sp}(r,t)$ vs. $r/\ell(t)$, for $\sigma = 0.6,0.8,1.6,\infty$. For the GD case in Fig.~\ref{f3}(a), we define the length scale $\ell (t)$ as the distance over which $C_{\rm sp}(r,t)$ decays to 0.2 of its maximum value $C_{\rm sp}(0,t) = 1$. For the KD case in Fig.~\ref{f3}(b), $C_{\rm sp}(r,t)$ exhibits oscillations due to the conservation law which dictates $\int d\vec{r}~C_{\rm sp}(\vec{r},t) = 0$. In this case, $\ell(t)$ is calculated from the first zero crossing. The spatial correlation function obeys dynamical scaling, as was shown in earlier work. In Figs.~\ref{f3}(a)-(b), we test for {\it super universality} (SU) by plotting $C_{\rm sp}(r,t)$ vs. $r/\ell(t)$ at fixed $t$ for different values of $\sigma$. The neat data collapse shows that $C_{\rm sp}$ obeys SU, i.e., the $\sigma$-dependence of the morphology is captured entirely by $\ell (t)$.

In Figs.~\ref{f3}(c)-(d), we show data for the autocorrelation function $C(t,t_w)$. We have confirmed that $C(t,t_w)$ obeys the scaling in Eq.~(\ref{autodecay1}) for different values of $\sigma$ (not shown here). In Fig.~\ref{f3}(c), we plot the GD scaling functions for $\sigma = 0.6, 0.8, 1.6, \infty$ at $t_w = 32$. The scaling functions clearly depend on $\sigma$ and do not obey SU. This is also true for KD, as shown in Fig.~\ref{f3}(d) for the same values of $\sigma$ at $t_w = 1024$.

\begin{figure}[htb]
\centering
\includegraphics[width=0.9\linewidth]{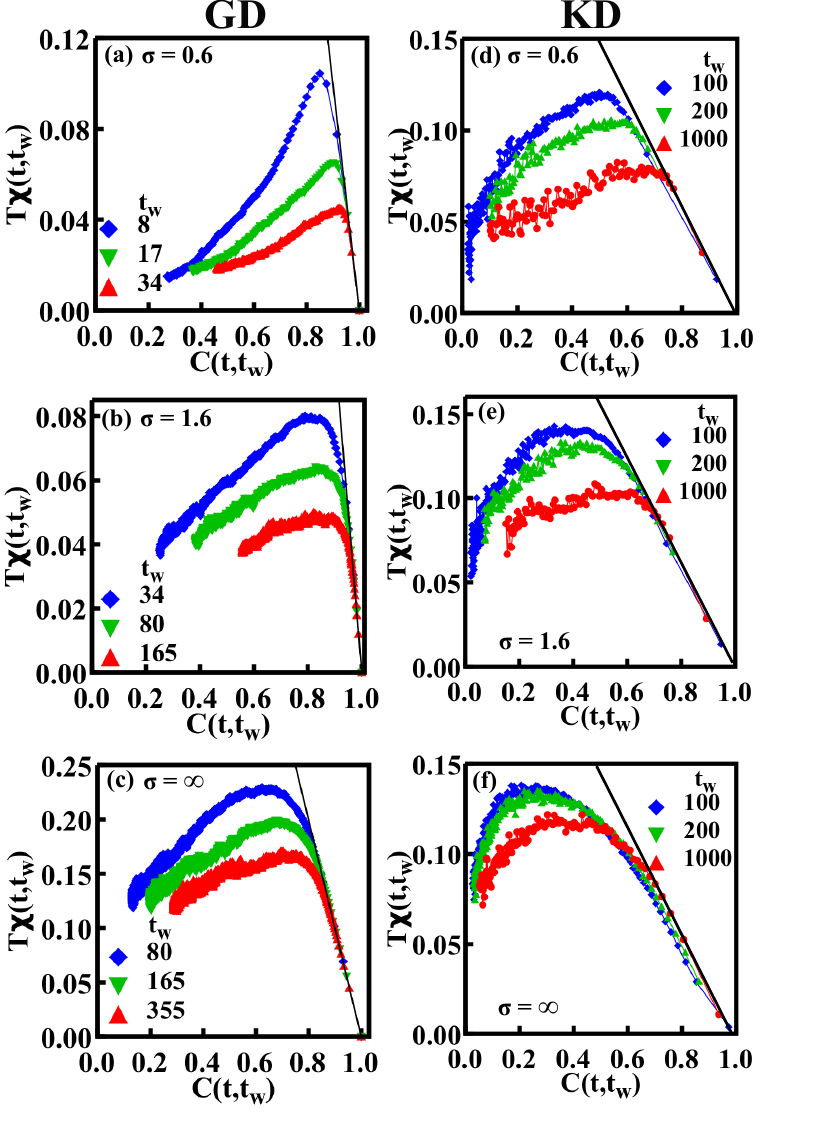}
\caption{Plot of $T \chi(t,t_w)$ vs. $C(t,t_w)$ for the LRIM. We plot data for 3 values of $t_w$ (as indicated) for (a) GD with $\sigma=0.6$, (b) GD with $\sigma=1.6$, (c) GD with $\sigma=\infty$, (d) KD with $\sigma=0.6$, (e) KD with $\sigma=1.6$, (f) KD with $\sigma=\infty$. The straight line in each frame denotes the FDT line.}
\label{f4}
\end{figure}

We next address the primary focus of our work, namely FDT violations in the LRIM. Fig.~\ref{f4} shows the plot of $T \chi(t,t_w)$ [obtained using Eq.~(\ref{RF_ij})-(\ref{IRF_av})] vs. $C(t,t_w)$ [obtained using Eq.~(\ref{AutoCor_N})]. The left and right panels correspond to GD and KD, respectively. We show data for $\sigma = 0.6, 1.6, \infty$ at 3 well-spaced values of $t_w$. The solid line in each frame is the FDT line represented by Eq.~(\ref{FDT_line}). Some important observations regarding Fig.~\ref{f4} are as follows.

For early times ($t \gtrsim t_w$), the LRIM with GD and KD obeys the FDT, which indicates thermodynamic equilibrium with the heat bath in this time regime. The dynamics in this regime is dominated by thermal fluctuations in the bulk domains, corresponding to the equilibrium behavior. As our simulations are performed at low $T$, the bulk fluctuations are not pronounced -- hence the FDT regime is relatively short-lived in Fig.~\ref{f4}. Furthermore, this short FDT regime is shortest for small values of $\sigma$, where domain growth is fastest and the aging regime manifests earlier. Data for $t \gg t_w$ show strong FDT violations, suggesting that the system retains memory over a long time. This aging regime is dominated by the slow dynamics of the interface motion. Based on the above discussion, we will interpret our numerical data for various two-time quantities as arising almost entirely from the aging regime.

Next, we study the time dependence and scaling behavior of $\chi (t,t_w)$ and $T_{\rm eff} (t,t_w)$. In the aging regime, the expected scaling behavior of $\chi (t,t_w)$ is \cite{mz09}
\begin{equation}
\chi(t,t_{w}) =  t_w^{-a} g\left( \frac{t}{t_{w}} \right) ,
\label{irfscal}
\end{equation}
where $a$ is a universal exponent. (We shall later refer to $a$ as the {\it susceptibility exponent}.) For the $d=2$ NN Ising model with GD, Lippiello et al. \cite{Lippiello2005} have verified the scaling form in Eq.~(\ref{irfscal}) with $a \simeq 0.26$.

\begin{figure}[htb]
\centering
\includegraphics[width=0.75\linewidth]{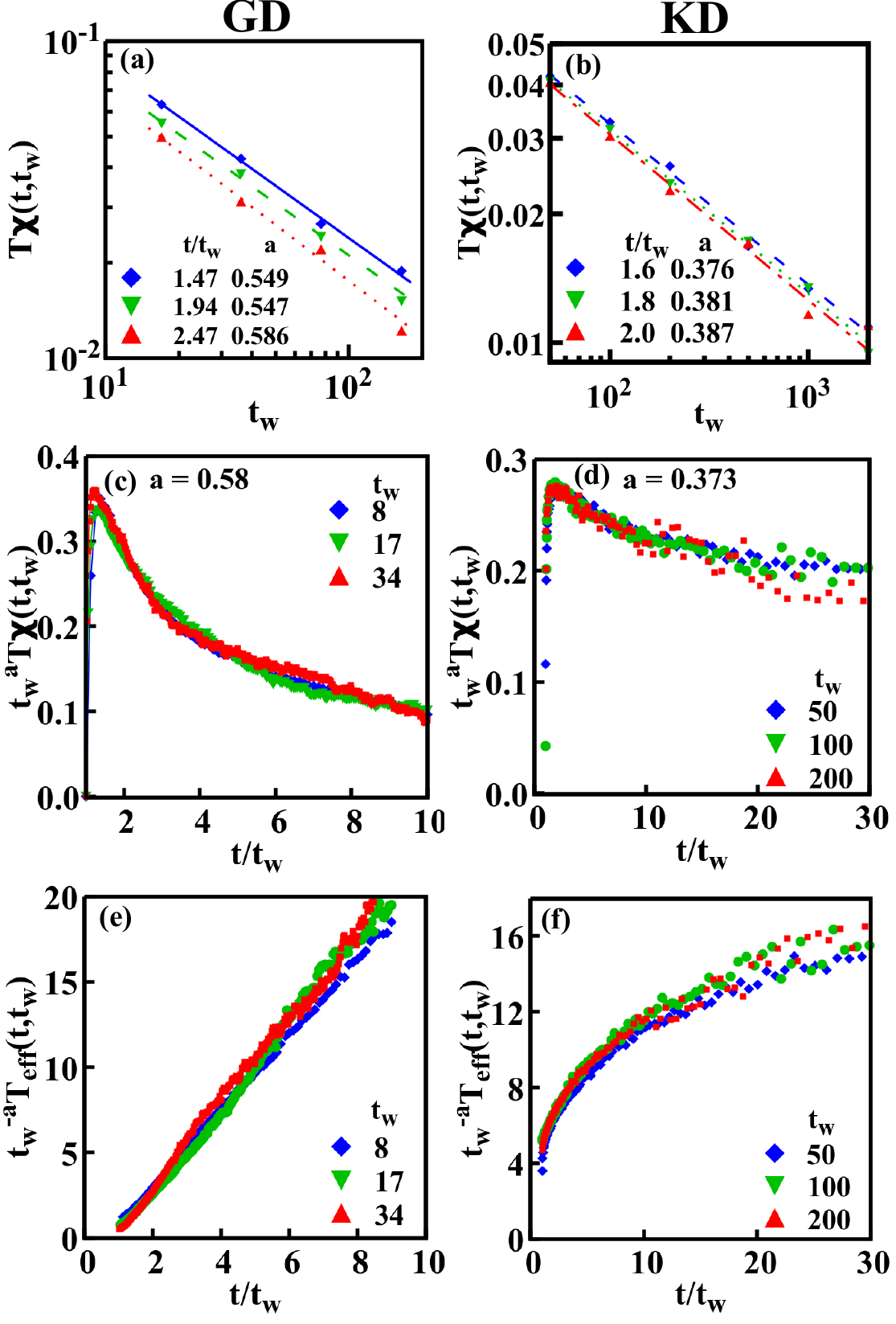}
\caption{Plots of $T \chi(t,t_w)$ vs. $t_w$ on a log-log scale for (a) GD with $\sigma = 0.6$, (b) KD with $\sigma = 0.6$. The data sets correspond to fixed values of $t/t_w$, as indicated. The best-fit lines yield estimates of $a$. Scaling plots of $t_w^a T \chi(t,t_w)$ vs. $t/t_w$ with 3 values of $t_w$ for (c) GD with $\sigma = 0.6$, (d) KD with $\sigma = 0.6$. Scaling plots of $t_w^{-a} T_{\rm eff} (t,t_w)$ vs. $t/t_w$ with 3 values of $t_w$ for (e) GD with $\sigma = 0.6$, (f) KD with $\sigma = 0.6$.}
\label{f5}
\end{figure}

In Fig.~\ref{f5}(a), we plot $T \chi(t,t_{w})$ vs. $t_w$ for the GD-LRIM with $\sigma = 0.6$. We plot data on a log-log scale for 3 fixed values of $t/t_w$, specified in the frame. These data are fitted with straight lines to obtain $a$ -- the corresponding values are specified in Fig.~\ref{f5}(a). Together with estimates of $a$ from other fixed values of $t/t_w$, we obtain an average of $a \simeq 0.58$. In Fig.~\ref{f5}(c), we plot data for $t_w^a T \chi(t,t_{w})$ vs. $t/t_w$ for 3 values of $t_w$ and $a = 0.58$. The neat data collapse confirms the scaling form in Eq.~(\ref{irfscal}). In Fig.~\ref{f5}(b), we ascertain $a$ for the KD case with $\sigma = 0.6$. The average value of $a$ for the KD-LRIM is $a \simeq 0.38$. The corresponding data collapse is shown in Fig.~\ref{f5}(d). The data collapse is reasonable, though there is some scatter at large values of $t/t_w$.

What is the scaling behavior of $T_{\rm eff} (t,t_w)$? We recall the definition from Eqs.~(\ref{FDR})-(\ref{GFDT}):
\begin{equation}
T_{\rm eff} (t,t_w) = \frac{T}{X(t,t_w)} = \frac{\partial_{t_w} C(t,t_w)}{R(t,t_w)}.
\label{teff}
\end{equation}
We know $C(t,t_w) = h(t/t_w)$, which yields $\partial_{t_w} C(t,t_w) h(t/t_w)~(-t/t_w^2)$. In addition, $R(t,t_w) = - \partial_{t_w} \chi (t,t_w)$, yielding $R(t,t_w) = t_w^{-1-a} p(t/t_w)$. We substitute these into Eq.~(\ref{teff}) to obtain
\begin{equation}
T_{\rm eff} (t,t_w) = t_w^a q(t/t_w) ,
\label{Tscal}
\end{equation}
where $q(x)$ is the scaling function. In Fig.~\ref{f5}(e), we superpose data for $t_w^{-a} T_{\rm eff} (t,t_w)$ vs. $t/t_w$ for GD with 3 values of $t_w$ and $\sigma = 0.6$. Although there is some scatter at late times, the data shows a reasonable collapse, confirming the above scaling function. We point out that even better scaling arises if the exponent $a$ in Eq.~(\ref{Tscal}) is allowed to be independent of the susceptibility exponent in Eq.~(\ref{irfscal}). However, this is not appealing because it requires a more complicated scaling framework.

Notice that $T_{\rm eff} \rightarrow \infty$ as $t \rightarrow \infty$ \cite{Barrat1998,marco2007}. This is because coarsening systems never settle to equilibrium in infinite systems, and domain growth proceeds forever. (Any saturation of $T_{\rm eff}$ is a finite-size effect that can cause freezing of domain growth at late times.) In Fig.~\ref{f5}(f), we show the corresponding scaling plot of $T_{\rm eff}$ for the KD-LRIM with $\sigma = 0.6$. Again, the data show a reasonable collapse, and $T_{\rm eff}$ diverges as $t \rightarrow \infty$. As expected, the divergence is slower than for the GD case.

\begin{figure}[htb]
\centering
\includegraphics[width=0.9\linewidth]{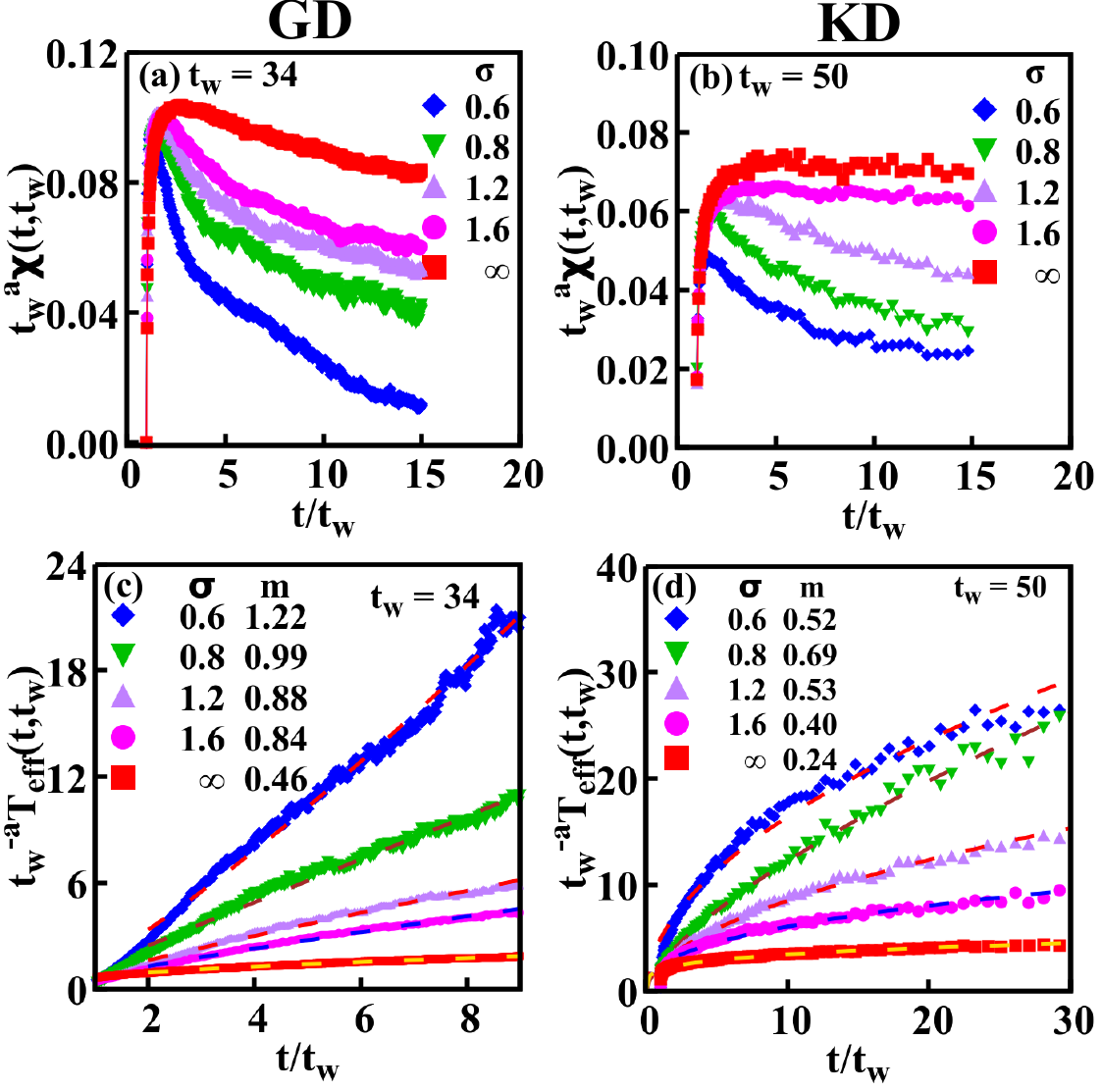}
\caption{The upper row shows plots of $t_w^a \chi(t,t_w)$ vs. $t/t_w$ for the indicated values of $\sigma$. We show data for (a) GD with $t_w=34$, (b) KD with $t_w=50$. The lower row shows plots of $t_w^{-a} T_{\rm eff} (t,t_w)$ vs. $t/t_w$. We show data for (c) GD with $t_w=34$, (b) KD with $t_w=50$. In (c) and (d), we also show power-law fits to the data as dashed lines. The values of the exponent $m$ are specified in the frames.}
\label{f6}
\end{figure}

In Fig.~\ref{f6}, we plot the scaled data for $\chi$ and $T_{\rm eff}$ for a single value of $t_w$ and for different values of $\sigma$. The frames on the left and right correspond to GD and KD, respectively. The scaling functions have a strong dependence on $\sigma$, showing the expected absence of SU, as for $C(t,t_w)$ in Figs.~\ref{f3}(c)-(d). We have also examined the plots of $t_w^a \chi(t,t_w)$ vs. $\ell(t)/\ell(t_w)$ and $t_w^{-a} T_{\rm eff} (t,t_w)$ vs. $\ell(t)/\ell(t_w)$. These also confirm the absence of SU, though we do not show these here for the sake of brevity. In Fig.~\ref{f6}(c), we attempt power-law fits ($y \sim x^m$) to the scaling functions of $T_{\rm eff}$ for GD. The data is well-described by power laws, though the exponents $m$ (specified in the frames) have a strong dependence on $\sigma$. In Fig.~\ref{f6}(d), we show that the KD data for $T_{\rm eff}$ also shows a power-law divergence, although the exponents are lower than those for GD.

\begin{figure}[htb]
\centering
\includegraphics[width=0.9\linewidth]{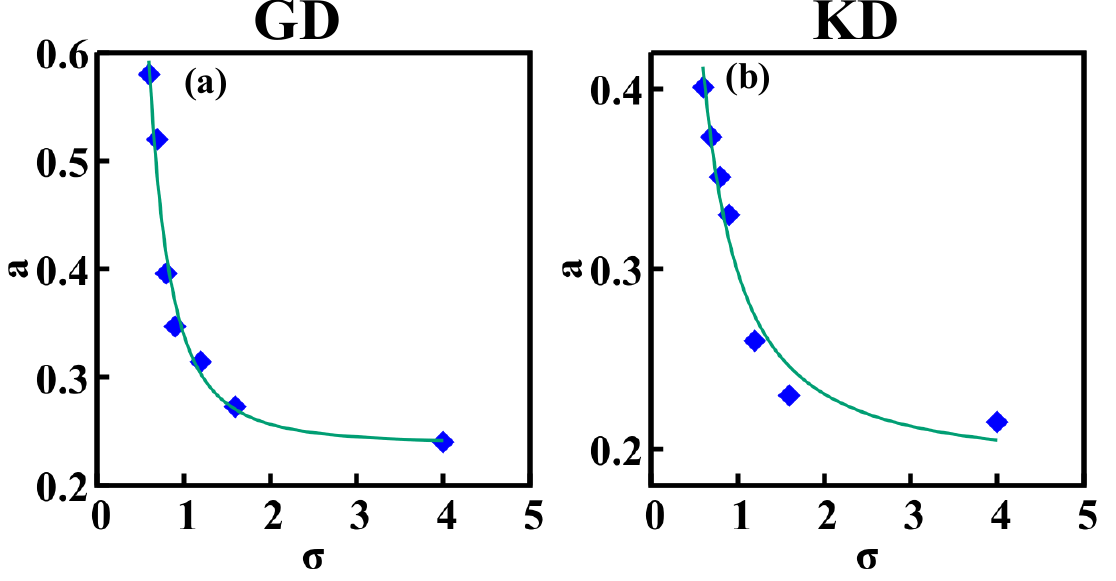}
\caption{(a) Plot of susceptibility exponent $a$ vs. $\sigma$ for GD. The solid line is a smooth interpolation of the data. (b) Analogous to (a) but for KD.}
\label{f7}
\end{figure}

Finally, in Fig.~\ref{f7}, we plot $a$ vs. $\sigma$ for GD (Fig.~\ref{f7}(a)) and KD (Fig.~\ref{f7}(b)). In contrast to the growth exponent $z$, the susceptibility exponent $a$ does not show a kink at $\sigma = 1$. Rather, it shows a smooth monotonic dependence on $\sigma$ with an asymptotic value of $a \simeq 0.25$ for $\sigma = \infty$. This is consistent with the NN result of Lippiello et al. \cite{Lippiello2005}. For KD, the corresponding value for $\sigma = \infty$ is $a \simeq 0.21$.

\section{Summary and Discussion}
\label{s4}

Let us conclude by summarizing our results regarding the violation of the FDT in the LRIM. In this context, let us recall the physical interpretation of the effective temperature $T_{\rm eff}$ \cite{cugliandolo2011effective}. It has been introduced to describe the deviation from equilibrium in a manner that still retains a thermodynamic interpretation. For fast fluctuations at an early time $t \gtrsim t_w$, the system behaves as if it is in local equilibrium with the heat bath, that is, $T_{\rm eff}= T$. In the late-time aging regime, when $t\gg t_w$, slow rearrangements or modes behave as if they were at higher temperature $T_{\rm eff} > T$. In addition, saturation of $T_{\rm eff}$ indicates that aging stops in certain degrees of freedom in that time sector, and the system behaves as if it has reached an equilibrium regime where fluctuations are governed by a thermal bath at $T_{\rm eff}$. A finite effective temperature allows for a quasi-thermodynamic description of an out-of-equilibrium system \cite{marco2001,Ludovic2022}. It allows us to define entropy, free energy, or even work relations for those slow degrees of freedom that have stopped aging. In {\it coarsening systems}, as stated in Eq.~(\ref{AC_Ag}), there is a two-time scale separation: a fast equilibrated regime arising from thermal excitations of the spins in the bulk domains; and a slow aging regime, where correlations decay slowly. Physically, this means that the system has large effective fluctuations, much larger than those at equilibrium. These large fluctuations arise from the movement of domain walls rather than from thermal excitations. As the process of domain growth through the annihilation of interfacial defects continues indefinitely, no finite temperature can describe the dynamics of aging. Consequently, for coarsening systems, $T_{\rm eff} \rightarrow \infty$ as $t \rightarrow \infty$. 

In light of the above discussion, we arrive at the following conclusions. \\
(i) There are strong violations of the FDT in the LRIM. They depend on the range of spin-spin interactions and conservation laws. \\
(ii) In the LRIM with GD or KD, $T_{\rm eff}\rightarrow \infty$ with time, and the approach to $\infty$ is power-law in both $t$ and $\ell (t)$. The monotonic evolution suggests that the system continuously explores the free-energy landscape with (local) rearrangement of the spins. This process becomes harder and harder as the system ages, due to the annihilation of interfacial defects. \\
(iii) In the aging regime, various two-time quantities such as $C(t,t_w), R(t,t_w), \chi (t,t_w)$ and $T_{\rm eff} (t,t_w)$ obey dynamical scaling for a fixed value of $\sigma$. However, they do not show superuniversal (SU) scaling, i.e., the scaling functions depend explicitly on $\sigma$. \\
(iv) The spatial correlation function $C_{\rm sp}(r,t)$ shows both dynamical scaling and SU. Thus, the domain morphology is independent of $\sigma$ -- the entire dependence on $\sigma$ is incorporated into the domain growth law. \\
(v) The domain growth law due to Bray and Rutenberg \cite{Bray1994} shows a kink at $\sigma = 1$, with $\ell(t) \sim t^{1/2}$ for GD and $\ell(t) \sim t^{1/3}$ for KD when $\sigma > 1$. The susceptibility exponent studied here does not show a kink and changes smoothly through $\sigma = 1$.

In conclusion, recent computational advances have enabled large-scale simulations of the LRIM. In this context, we investigate phase ordering kinetics in the $d=2$ LRIM through MC simulations with conserved and nonconserved dynamics. In recent years, there have been several studies of the spatial properties of the kinetic LRIM, e.g., $C_{\rm sp}(r,t), \ell(t)$. Some studies of two-time quantities have also been carried out such as $C(t,t_w)$. These have focused primarily on the decay exponent of $C(t,t_w)$. To the best of our knowledge, there has been no study of the FDT violation, response function, and susceptibility in this system. The present study addresses these lacunae. In spite of its simplicity, the LRIM provides some important lessons in understanding out-of-equilibrium dynamics by quantifying FDT violations. Although many aspects of the LRIM still remain unexplored, we hope that our investigations provide a comprehensive framework for studying related models of aging dynamics.

\subsection*{\bf Acknowledgments} VB thanks Marco Zannetti for useful discussions in the early part of this work. PS acknowledges UGC, India, for a research fellowship. PS and VB gratefully acknowledge the HPC facility of IIT Delhi for computational resources. VB acknowledges ANRF, India, for a CORE research project.

\newpage
\bibliographystyle{h-physrev.bst}
\bibliography{ref.bib}

\newpage

\end{document}